\documentclass[prd,twocolumn,superscriptaddress,floatfix,amsmath,amssymb,amsfonts,nofootinbib,longbibliography]{revtex4-2}

 \usepackage{amsmath}
\usepackage{graphicx}
\usepackage{epstopdf}
\usepackage{float}
\usepackage{hyperref}
\usepackage{color}
\usepackage[T1]{fontenc}
\usepackage[utf8]{inputenc}
\usepackage[toc,page]{appendix}
\usepackage[usenames,dvipsnames]{xcolor}
\usepackage[normalem]{ulem}
\usepackage{lipsum, babel}
\usepackage{natbib}

\newcommand{\be}{\begin{equation}}
\newcommand{\ee}{\end{equation}}
\newcommand{\ba}{\begin{eqnarray}}
\newcommand{\ea}{\end{eqnarray}}

\newcommand{\beq}{\begin{equation}}
\newcommand{\eeq}{\end{equation}}
\newcommand{\beqa}{\begin{eqnarray}}
\newcommand{\eeqa}{\end{eqnarray}}

\begin{document}

\title{Migrating Carrollian Particles on Magnetized Black Hole Horizons}

\author{Ji{\v r}{\'i} Bi{\v c}{\'a}k}
\email{bicak@mbox.troja.mff.cuni.cz}
\affiliation{Institute of Theoretical Physics, Faculty of Mathematics and Physics,
Charles University, Prague, V Hole\v{s}ovi\v{c}k\'ach 2, 180 00 Prague 8, Czech Republic}

\author{David Kubiz\v{n}{\'a}k}
\email{david.kubiznak@matfyz.cuni.cz}
\affiliation{Institute of Theoretical Physics, Faculty of Mathematics and Physics,
Charles University, Prague, V Hole\v{s}ovi\v{c}k\'ach 2, 180 00 Prague 8, Czech Republic}

\author{T. Rick Perche}
\email{trickperche@perimeterinstitute.ca}

\affiliation{Department of Applied Mathematics, University of Waterloo, Waterloo, Ontario, N2L 3G1, Canada}
\affiliation{Perimeter Institute for Theoretical Physics, Waterloo, Ontario, N2L 2Y5, Canada}
\affiliation{Institute for Quantum Computing, University of Waterloo, Waterloo, Ontario, N2L 3G1, Canada}

\begin{abstract}
By considering a misaligned (asymptotically uniform) magnetic field in the background of a rotating black hole, we uncover a possibility for a highly non-trivial  motion of Carrollian particles on the black hole horizon that is 
characterized by a time-dependent velocity field and reminds us (because of its latitudinal oscillations) of a  
`monarch butterfly migration'.
\end{abstract}

\date{February 22, 2023}


\maketitle

\section{Introduction}
{\em Carrollian particles} are massless particles that live on a null surface equipped with a Carroll structure, a horizon of a black hole, for example. 
Contrary to Carroll fluids that are known to feature non-trivial dynamics, e.g. \cite{Bergshoeff:2014jla,Freidel:2022bai,Freidel:2022vjq, Bagchi:2023ysc}, it was believed that a single Carroll particle cannot move \cite{VD1966,Bergshoeff:2014jla,Duval:2014uva}. However, as  
recently shown this is no longer true in $(2+1)$-dimensional Carrollian settings in the presence of magnetic fields \cite{Marsot:2022qkx, Marsot:2022imf}. Namely, considering a {\em double central extension} of the Carroll group, the massless Carroll particle with anyonic spin can move under the influence of a {\em magnetic field} $B$, according to the following equation:
\begin{equation}\label{eq:EoM}
    \frac{dx^A}{dv} =  \frac{\mu \chi}{\kappa_{\text{mag}}} \epsilon^{AB}\partial_B B\,,
\end{equation}
where $v$ is the (preferred on the horizon) Carrollian time. 
Parametrizing the 2-dimensional particle's trajectory $x^A=x^A(v)$, $A=1,2$, $\mu$ is the magnetic moment of the particle, $\chi$ its anyonic spin, $\kappa_{\text{mag}}$ is a central extension parameter which allows for the particle to couple to  electromagnetism, and $\epsilon^{AB}$ is the 2-dimensional Levi-Civita tensor of the 2-dimensional metric on the horizon (including its determinant).

For its remarkable similarity with the  
following (effectively 2-dimensional) 3D equation: 
\begin{equation} \label{eq:analSpinHall}
    \frac{dx^i}{dv} = -e (E \times \Theta)^i\,,
\end{equation}
of the spin-Hall effect, e.g. \cite{hirsch1999, Harte:2022dpo} 
(where $e$ is the electric charge of the spinning particle, $\Theta$ its Berry curvature, and $E$ the external electric field), the motion described by Eq.~\eqref{eq:EoM} was dubbed the 
{\em anyonic spin-Hall effect} \cite{Marsot:2022qkx, Marsot:2022imf}. 

Up to now, the  anyonic spin-Hall effect was studied for the magnetized horizons of the Kerr--Newmann solution (endowed possibly with a magnetic monopole) \cite{Marsot:2022qkx, Marsot:2022imf, Gray:2022svz}, or for the Kerr black hole immersed in an aligned test uniform magnetic field \cite{Gray:2022svz}. However, in all these cases, the induced motion of Carrollian particles is sort of `trivial' -- the particles orbit the black hole horizon on `circular trajectories' around the axis of rotation of the black hole.

The goal of the present paper is to show that a much more interesting motion of Carrollian particles is possible. To this purpose, we employ a remarkable test field in the vicinity of a rotating (Kerr) black hole, that represents an (asymptotically uniform) magnetic field {whose asymptotic axis of symmetry is tilted with respect to the rotational axis of the hole}
\cite{bivcak1976stationary, bivcak1985magnetic}. {Since a stationary black hole must be either static or axisymmetric \cite{hawking1972black}, a Kerr hole immersed in a non-axisymmetric perturbing field evolves in time -- either it loses its angular momentum or its axis becomes aligned with respect to the perturbing field. For the first analysis of this effect, using scalar field but indicating also the effect for gravitational perturbing field, see \cite{press1972time}. The case of the external non-aligned magnetic field is quite comprehensibly analysed in \cite{thorne1986black}.  There is a number of other works, related also to the so-called Bardeen-Petterson effect, with qualitatively similar results: the component of the hole's angular momentum orthogonal to the external field decays exponentially with the e-folding time extremely large. For the case of the magnetic field, \cite{thorne1986black} determine the e-folding time to be  $10^{10} \, \mbox{years}\, (10^4 \mbox{Gauss}/B)^2 \, (10^8 M_{\odot}/M)$.}
In astrophysical situations, such a field is produced by the magnetized plasma accreting onto the black hole, with an accretion disc misaligned with the rotation axis \cite{king1977magnetic}. In \cite{thorne1986black} the astrophysical applications of black hole electrodynamics are summarized in Chapter IV. For more recent references see, for example, \cite{Kim:2002ei, McKinney:2012wd, Liska:2018ayk}.
As we shall see, the tilted magnetic field induces a rather curious 
motion of Carrollian particles. This motion  is characterized by a time-dependent velocity vector field and, due to its dependence on the latitudinal coordinate, it is analogous to the `monarch butterfly migration', e.g. \cite{reppert2018demystifying}. 

{The remaining part of the paper is organized as follows.} 
In Sec.~\ref{sec2} we review the misaligned magnetic field. Sec.~\ref{sec3} describes the corresponding Carrollian structure induced in the horizon. In Sec.~\ref{sec4} we analyze the motion of Carrollian particles on the horizon in the presence of the tilted magnetic field. The conclusions of our work can be found in Sec.~\ref{sec5}.

\section{Rotating black hole immersed in  a misaligned magnetic field}\label{sec2}
A rotating black hole is described by the Kerr metric, which in the standard Boyer--Lindquist coordinates reads
\ba
ds^2&=&-\frac{\Delta}{\Sigma}\bigl(dt-a\sin^2\!\theta d\phi\bigr)^2+\frac{\sin^2\!\theta}{\Sigma}\bigl(adt-(r^2+a^2)d\phi\bigr)^2\nonumber\\
&&+\frac{\Sigma}{\Delta}dr^2+\Sigma d\theta^2\,,
\ea
where 
\be
\Delta=r^2-2Mr+a^2\,,\quad \Sigma=r^2+a^2\cos^2\!\theta\,. 
\ee
In these expressions, $M$ stands for the mass of the hole and $J=Ma$ is its angular momentum around the axis of symmetry -- the `$z$-axis'.

The {(outer) {\em black hole horizon}} is located at the largest root of $\Delta(r_+)=0$, at 
\be
r_+=M+\sqrt{M^2-a^2}\,. 
\ee
{The horizon is dragged due to the rotation of the hole, and rotates 
with the 
angular velocity:}
\be
\Omega_+=\frac{a}{r_+^2+a^2}\,,
\ee
equal to the angular velocity of zero angular momentum observers (ZAMOs) at the outer horizon.

When $a\to M$, the black hole is called {\em extremal}. It possesses a degenerate horizon and its Hawking temperature vanishes. In this limit, the aligned magnetic field is expelled from the horizon, which is known as the {\em black hole Meissner effect} \cite{bivcak1976stationary, bivcak1985magnetic, Chamblin:1998qm, Penna:2014aza, Bicak:2015lxa}. However, this is no longer true for the tilted magnetic field which remains to penetrate the horizon even in the extremal case \cite{bivcak1985magnetic}.  (See also \cite{Gregory:2013xca} for the case of 
a magnetic flux tube of a cosmic string 
painted on the extremal Kerr horizon, which features a penetration/expulsion phase transition depending on the ratio of the thickness of the string and the horizon radius.)

The {\em tilted test magnetic field} is described by the following vector potential \cite{bivcak1976stationary, bivcak1985magnetic}:
\ba\label{tilted}
A_t&=&\frac{B_1aM}{\Sigma}\sin\theta\cos\theta\bigl(r\cos\psi-a\sin\psi\bigr)\nonumber\\
&+&\frac{B_0 aMr}{\Sigma}(1+\cos^2\!\theta)-B_0a\,,\nonumber\\
A_r&=&-B_1(r-M)\cos\theta \sin\theta \sin\psi\,,\nonumber\\
A_\theta&=&-B_1a(r\sin^2\!\theta +M\cos^2\!\theta)\cos\psi\nonumber\\
&-&B_1\Bigl(r^2\cos^2\!\theta+(a^2-Mr) \cos 2\theta\Bigr)\sin\psi\,,\nonumber\\
A_\phi&=&B_0\sin^2\!\theta\Bigl(\frac{r^2+a^2}{2}-\frac{a^2Mr}{\Sigma}(1+\cos^2\!\theta)\Bigr)\nonumber\\
&-&\frac{B_1\sin 2\theta}{2}\Bigl(
\Delta \cos\psi
+\frac{(r^2\!+\!a^2)M}{\Sigma}(r\cos\psi-a\sin\psi)\Bigr)\,.
\nonumber\\
\ea
Here, $B_1$ denotes the field component perpendicular to the rotation axis,  while $B_0$ 
is the component aligned with the axis, and $\psi$ is the azimuthal coordinate in the Kerr ingoing coordinates:
\be\label{psicoord}
d\psi=d\phi+\frac{a}{\Delta}dr\,. 
\ee
We refer the reader to \cite{bivcak1976stationary, bivcak1985magnetic, thorne1986black, Karas:2012mp} for an analysis of the physical properties of this field and for an illustration of its dragging around the black hole.


\section{Carrollian structure and magnetized horizon}\label{sec3}

In what follows, we will describe the Carrollian structure in the horizon of the black hole. 
To this purpose, we first transfer to the ingoing    
coordinates $(v, r, \theta, \varphi)$ which co-rotate with the horizon, by 
\ba
d\phi&=&d\varphi +\Omega_+ dv-\frac{a}{\Delta}dr\,,\nonumber\\ 
dt&=&dv-\frac{r^2+a^2}{\Delta}dr\,.
\ea 
{These differ from the standard Kerr ingoing coordinates $(v, r, \theta, \psi)$ by the additional $\Omega_+dv$ term in the first expression, c.f. \eqref{psicoord}, which `eliminates' the rotation of the horizon.
Upon this coordinate transformation,} the metric takes the following explicit form:\footnote{Note that in contrast to \cite{Marsot:2022qkx}, we use ingoing rather than outgoing coordinates. Consequently, in our case the Carrollian particles move on the black hole horizon rather than the white hole horizon, as is the case in \cite{Marsot:2022qkx}.}  
\begin{align}\label{eq:KNmetric}
ds^2 = &- \frac{\Delta}{\Sigma}\left({\frac{\Sigma_+}{r_+^2+a^2}dv - \frac{\Sigma}{\Delta} dr} - a \sin^2\!\theta d\varphi\right)^2\nonumber\\& + \frac{\sin^2 \theta}{\Sigma}\left({\Omega_+(r_+^2-r^2)dv} - (r^2 + a^2) d\varphi\right)^2 \nonumber\\&+ \frac{\Sigma}{\Delta} dr^2+ \Sigma d\theta^2\,,
\end{align}
where we have introduced 
\be
\Sigma_+=r_+^2+a^2\cos^2\!\theta\,, 
\ee
while the new components of the vector field are now
\ba
A_v'&=&\Omega_+A_\phi+A_t\,,\nonumber\\
A_r'&=&A_r-\frac{a}{\Delta}A_\phi-\frac{a^2+r^2}{\Delta}A_t\,,\nonumber\\
A_\theta' &=&A_\theta\,,\quad A_\varphi'=A_\phi\,. 
\ea
{Note that, in these coordinates
$
\psi=\varphi+\Omega_+v+\mbox{const.}, 
$ 
where $\text{const.}$ denotes an integration constant that will be set to zero for the remainder of our analysis, so that we have 
\be
\psi=\varphi+\Omega_+v\,. 
\ee
It is precisely this $\psi$ which introduces the time dependence into our problem. It originates from the fact that the preferred Carrollian time $v$ is defined in the co-rotating with the horizon coordinate frame. 
}

{To see this, let us define the 2-dimensional metric on the horizon} and the associated Carrollian structure. The horizon is generated by a null Killing vector field, which in the new coordinates simply reads 
\be
\xi=\partial_v\,. 
\ee
The corresponding $v$-coordinate is the (preferred) Carrollian time on the horizon.\footnote{Had we chosen the standard Kerr {ingoing coordinates, as opposed to the above co-rotating ones, we would not obtain the preferred Carrollian time, as in that case we would have $\xi=\partial_v+\Omega_+\partial_\psi$, and the  horizon metric \eqref{q} would appear} as `rotating in the azimuthal direction'.}  We can also obtain a null normal to the horizon, $n_\mu$, which is orthogonal to $\partial_\theta$ and $\partial_r$, such that $n_\mu \xi^\mu = 1$. It reads
\begin{equation}
    n = dv -\frac{a(r_+^2 + a^2)}{\Sigma_+} \sin^2 \theta d\varphi\,.
\end{equation}
Using $\xi$ and $n$ one can define the projector into the horizon as
\begin{equation}\label{projectorq}
 q^\mu{}_{\nu} = \delta^\mu_\nu - \xi^\mu n_\nu - n^\mu \xi_\nu\,.    
\end{equation}
With both indices down, 
this represents a degenerate metric on a $(2+1)$-dimensional horizon surface: 
\begin{equation}\label{q}
    q =q_{AB}dx^Adx^B= \Sigma_+ d\theta^2+\frac{(r_+^2 + a^2 )^2\sin^2 \theta}{\Sigma
    _+} d\varphi^2\,,
\end{equation}
with $x^A=\{\theta, \varphi\}$.
The horizon $\mathcal{H}$ can thus be endowed with a Carrollian structure \cite{Donnay:2019jiz, Bergshoeff:2022eog}.
This structure is given by a fibre bundle $p$: $\mathcal{H}\to S$, where $S$ has the topology of $S^2$ and is just any constant $v$ slice,  $S = \mathcal{H}\rvert_{v = v_0}$. The projection is the usual projector constructed from the metric, and the surface $S$ is a Riemannian manifold, equipped with the metric \eqref{q}. The fibre bundle has a vertical vector field given by $\partial_{{v}}$, which generates the vertical space -- the ``time'' evolution along the Carrollian time $v$. 

To obtain the magnetic field $B$, relevant for the motion of Carrollian particles, \eqref{eq:EoM}, we
define \cite{Gray:2022svz}
\be
B=\frac{1}{2}\epsilon^{AB}\hat F_{AB}\,, 
\ee
where $\epsilon^{AB}$ is the Levi-Civita tensor associated with the metric $q$ \eqref{q}, 
{$\epsilon^{\theta\varphi}=1/\sqrt{\det q_{AB}}=1/[(r_+^2+a^2)\sin\theta],$}
and $\hat F_{AB}$ is the projection of the bulk electromagnetic field strength
\be
\hat F_{AB}=q^\mu{}_A q^\nu{}_B F_{\mu\nu}\,. 
\ee
In particular, for the tilted test field \eqref{tilted} we (upon using the co-rotating with horizon coordinates) obtain
\ba\label{Bfinal}
    B &=& \frac{B_1 \sin \theta }{2 r_+ \Sigma_+^2} \Bigl(2 r_+ (a^4\cos^2\theta - r_+^4) \cos \psi\nonumber\\
   &&+a \bigl[(r_+^2 - a^2)a^2\cos^2\!\theta + r_+^2 (3r_+^2 +a^2)\bigr]
   \sin \psi\Bigr)\nonumber\\
&&+\frac{B_0
   (a^4-r_+^4)\cos\theta}{\Sigma_+^2}\,.
\ea

As a quick check let us consider various limits of this expression. First, for $B_1=0$, we obtain the result for the aligned magnetic field 
\be
B_{\mbox{\tiny align}}=\frac{B_0
   (a^4-r_+^4)\cos\theta}{\Sigma_+^2}\,,
\ee
studied in \cite{Gray:2022svz}. Second, in the extremal limit $a\to M$, the contribution from $B_0$ vanishes, and we recover 
\be
B_{\mbox{\tiny extr.}}=\frac{B_1\sin\theta}{(1+\cos^2\!\theta)^2}\bigl(2\sin\psi-\sin^2\!\theta \cos\psi\bigr)\,, 
\ee 
where $\psi=\varphi+v/(2a)$. Finally, in the Schwarzschild limit, $a\to 0$, we recover
\be
B_{\mbox{\tiny Sch}}=-B_0\cos\theta-B_1\sin\theta \cos\varphi\,. 
\ee
This is nothing but the uniform magnetic field around (spherically symmetric) Schwarzschild, written now in tilted coordinates.


\section{Carrollian motion in tilted magnetic field}\label{sec4}

\begin{figure}[h!]
\centering
    \includegraphics[
    width=0.49\textwidth]{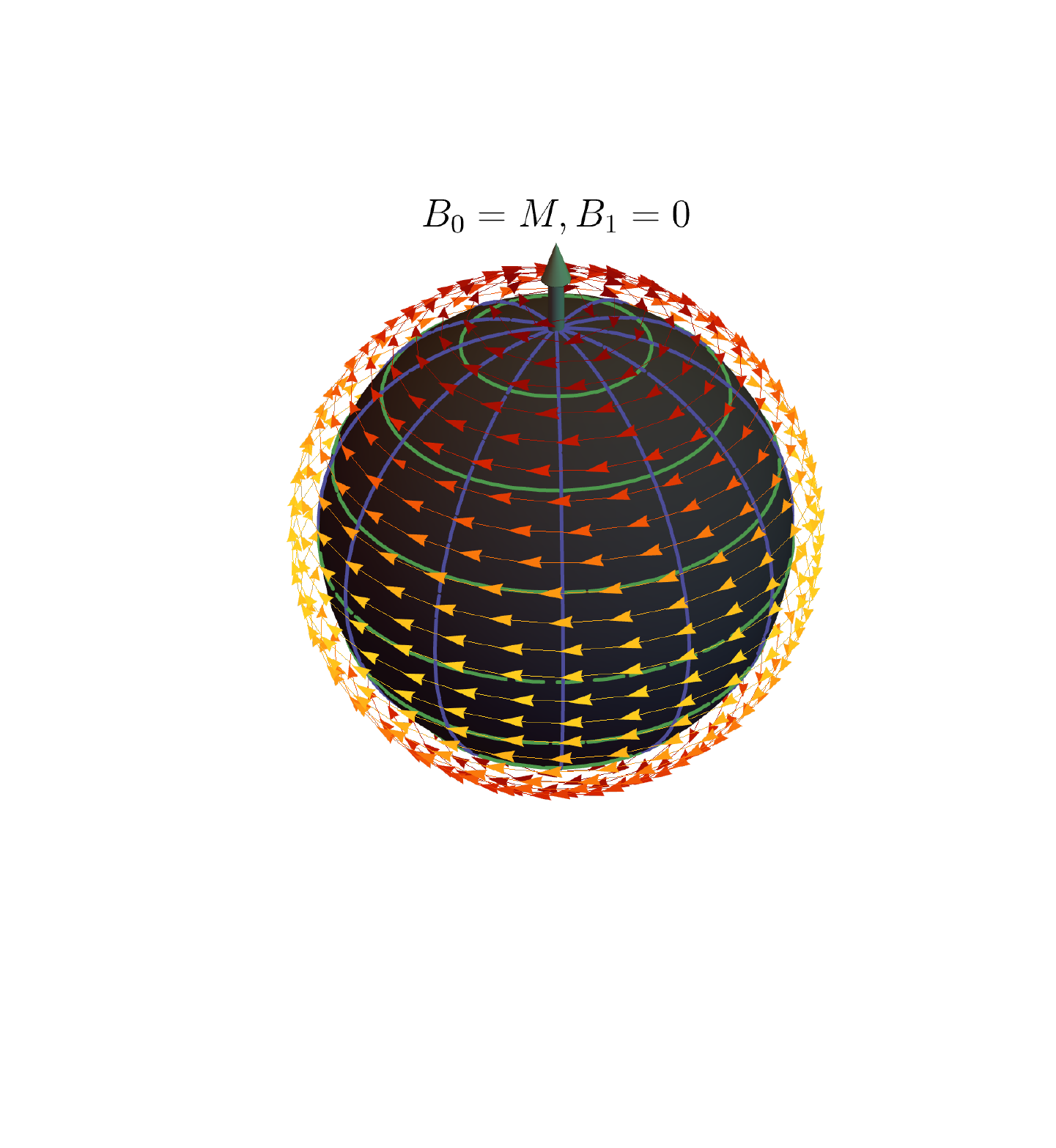}
\vspace{-4cm} 
   \newline
    \includegraphics[
    width=0.49\textwidth]{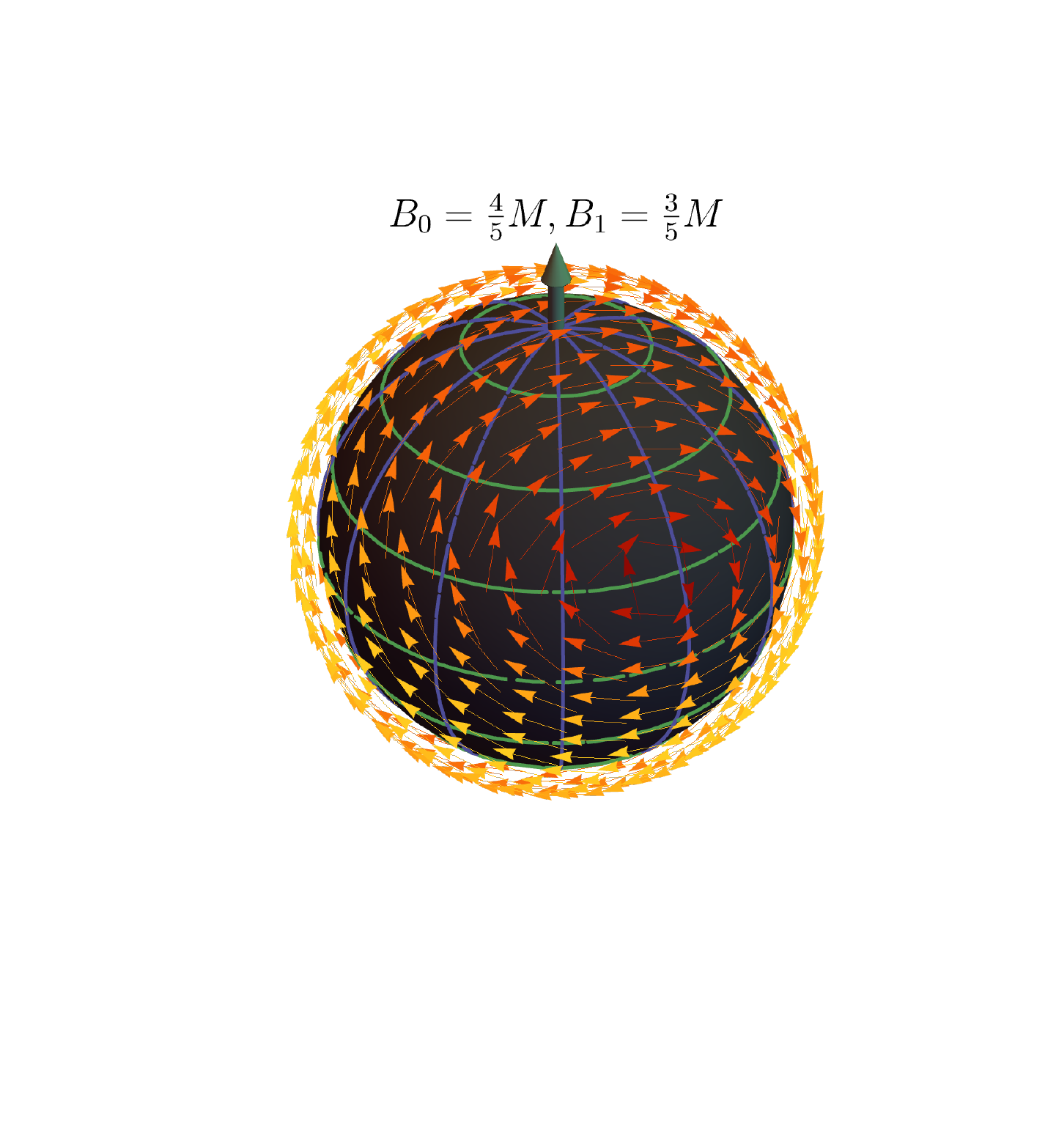}
\vspace{-4cm}
\newline    
    \includegraphics[
    width=0.49\textwidth]{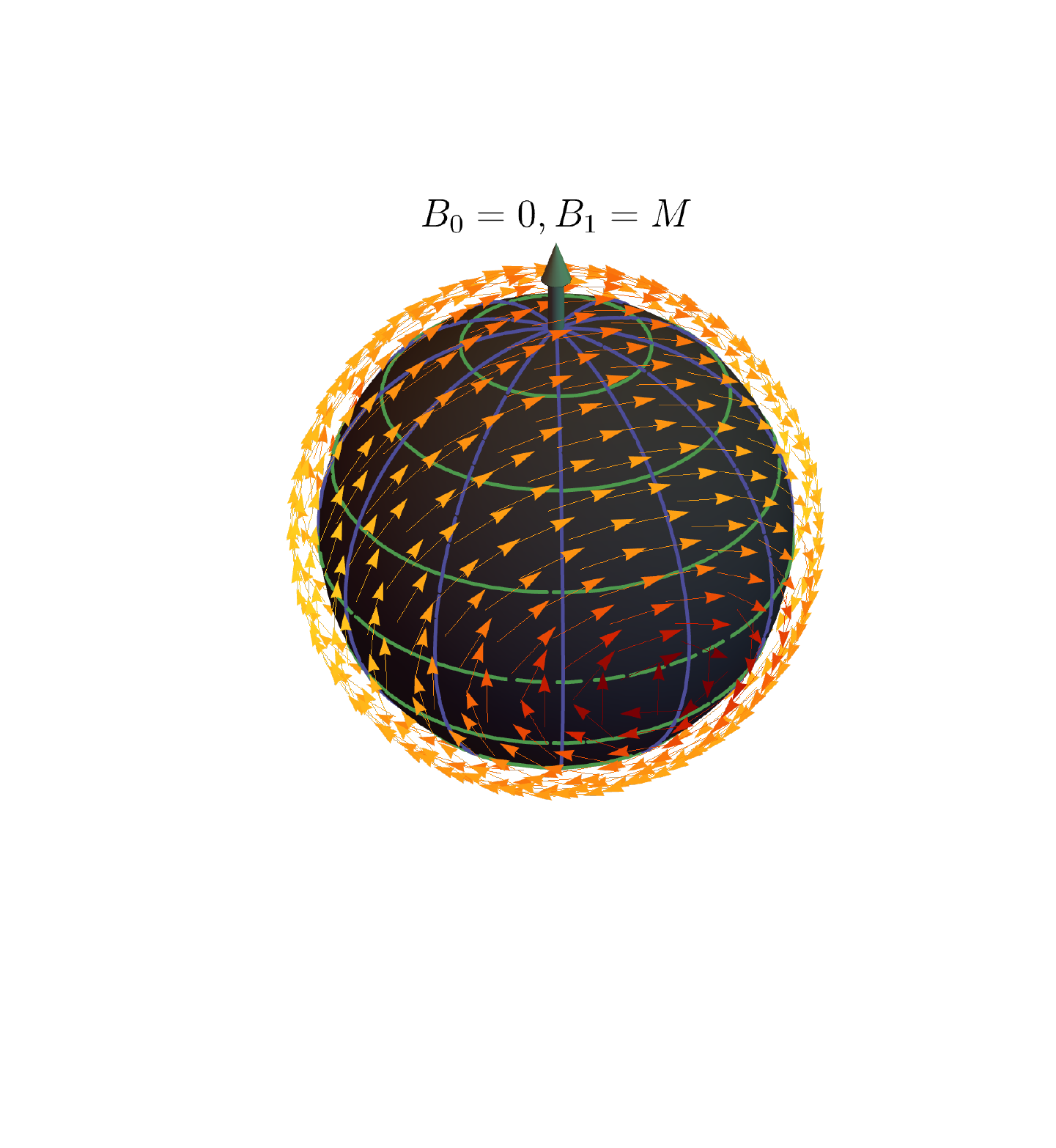}
\vspace{-3cm}
\caption{{\bf Velocity fields: field alignment.} We display velocity fields at $v=0$ for different values of $B_0$ and $B_1$ for the non-extremal rotating Kerr black hole characterized by $r_+=\frac{8}{5}M$ and $a=\frac{4}{5}M$. The large green arrow denotes the axes of rotation of the black hole, and the $\theta$ and $\varphi$ coordinate lines are depicted in green and blue, respectively. The arrows for the vector indicate its magnitude: red stands for smaller vectors and yellow for larger ones. } 
\label{fig:vel1}
\end{figure}

\begin{figure}[h!]
\centering
    \includegraphics[
    width=0.49\textwidth]{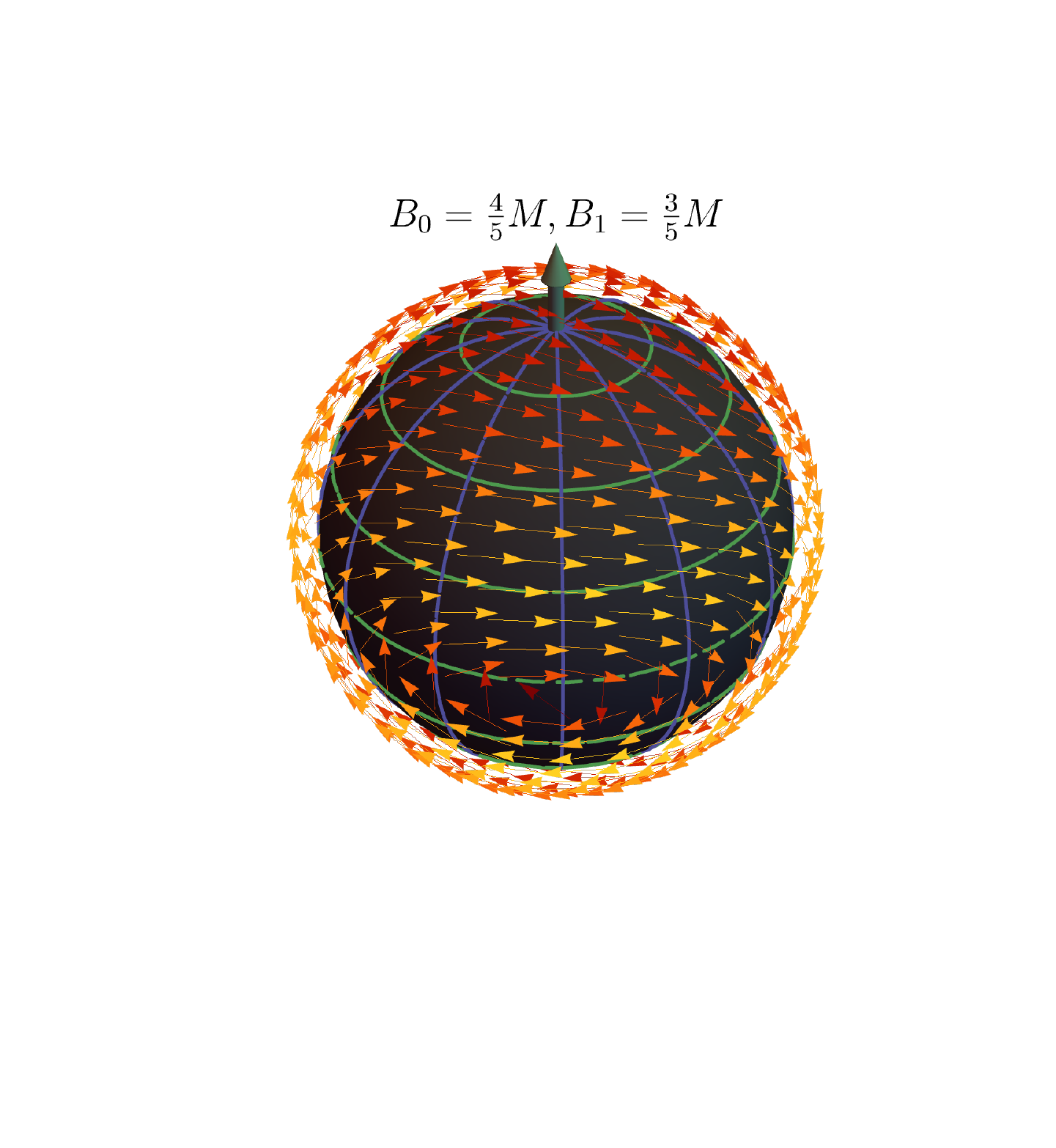}
\vspace{-4cm} 
   \newline
    \includegraphics[
    width=0.49\textwidth]{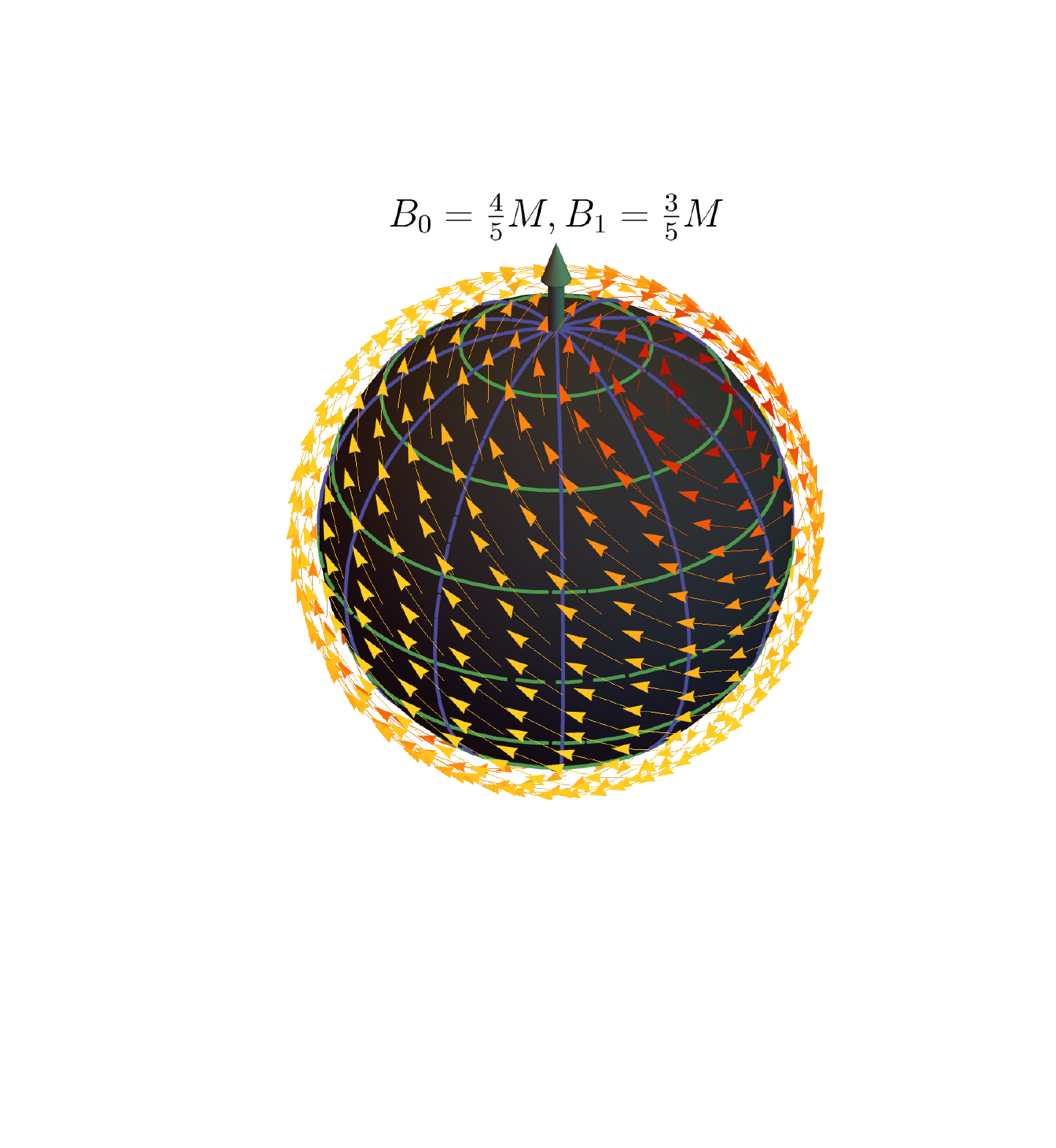}
\vspace{-3cm}
\caption{{\bf Velocity fields: effect of rotation.} We display the velocity field at $v=0$ for a fixed ratio of $B_0=\frac{4}{5}M$ and $B_1=\frac{3}{5}M$ for i) the extremal black hole (up) with $r_+=a=M$ and ii) non-rotating black hole with $r_+=2M$ and $a=0$. } 
\label{fig:vel2}
\end{figure}

In this section we analyze the Carrollian motion induced by the tilted magnetic field, and compare our results with the motion when the magnetic field is parallel to the axis of rotation of the black hole.

{
Let us start by analyzing the {velocity field 
\be 
\Bigl(\frac{d\theta}{dv},\frac{d\varphi}{dv}\Bigr)
\ee
induced by the magnetic field \eqref{Bfinal} via  Eq. \eqref{eq:EoM}. This is displayed} for $v=0$ and various ratios of $B_0$ vs. $B_1$ for the rotating non-extremal Kerr black hole in Fig.~\ref{fig:vel1}. The two limiting cases of the extremal black hole and a non-rotating black hole are for fixed ratio of $B_0$ and $B_1$ displayed in Fig.~\ref{fig:vel2}.\footnote{
The figures are meant for illustrative purposes only. The horizon is displayed as `spherical' and the $(\theta, \varphi)$ coordinates are identified with spherical angles. In particular, we ignore the fact that close to the polar regions, the horizon of a rotating black hole cannot be embedded in $\mathbb{R}^3$, e.g. \cite{Frolov:2006yb}.} 
Here, the red vectors denote the region where the norm of the velocity field is small, while yellow vectors denote large velocities.   We notice that there are two vortices (centers of red whirlwinds) on opposite sides of the black hole where the velocity field vanishes and 
around which the Carrollian particles revolve.

In the case where only $B_0$ is present, {the magnetic field is parallel to the axis of rotation and} the Carrollian particles revolve around this axis, as previously analyzed in~\cite{Gray:2022svz}. When $B_1$ is non-zero, the axis around which the Carrollian particles rotate shifts and gets positioned in an intermediary axis between the directions of $B_0$ and $B_1$. Also notice that the motion of the particles in this case is not a perfect circle around the axis. When $B_0 = 0$, we find that the coordinates of the points at which the velocity field vanishes at $v = 0$ are located {at $ \theta = \pi/2$ and $\tan\phi = - a(a^2+3 r_+^2)/(2r_+^3)$, which yields two solutions for $\phi$, which differ by $\pi$.} That is, the two points are antipodal, and define the axis {around which the Carrollian} particles orbit.

{Interestingly, the displayed velocity fields resemble the ``eddy currents'' around rotating black holes studied in the membrane paradigm in \cite{thorne1986black}. It would be interesting to probe if there is any deeper reason for this similarity.}

So far we have only analyzed the velocity field at $v = 0$. {As $v$ varies, both the magnetic field and the velocity fields 
precess around the axis of rotation of the  black hole with angular velocity $\Omega_+$, as
$\varphi \rightarrow \varphi + \Omega_+ v$.} 
{This precession is caused by the dragging of the hole.} Namely, while the {profile of the magnetic field outside the horizon remains static with respect to observers located at infinity, the horizon itself rotates with respect to these observers with angular velocity $\Omega_+$, and the field seen as constant at infinity will be seen as rotating by a particle at the horizon, as is the case of the Carrollian particles discussed here.}\footnote{Let us remark, however, that while the profile of the field {remains stationary outside the hole ergosphere, it is only uniform at infinity, and becomes more and more dragged along the rotation of the hole as we approach the horizon, see \cite{Karas:2012mp} for illustrations of this fact.}}

\begin{figure}[h!]
\centering
    \includegraphics[
    width=0.4\textwidth]{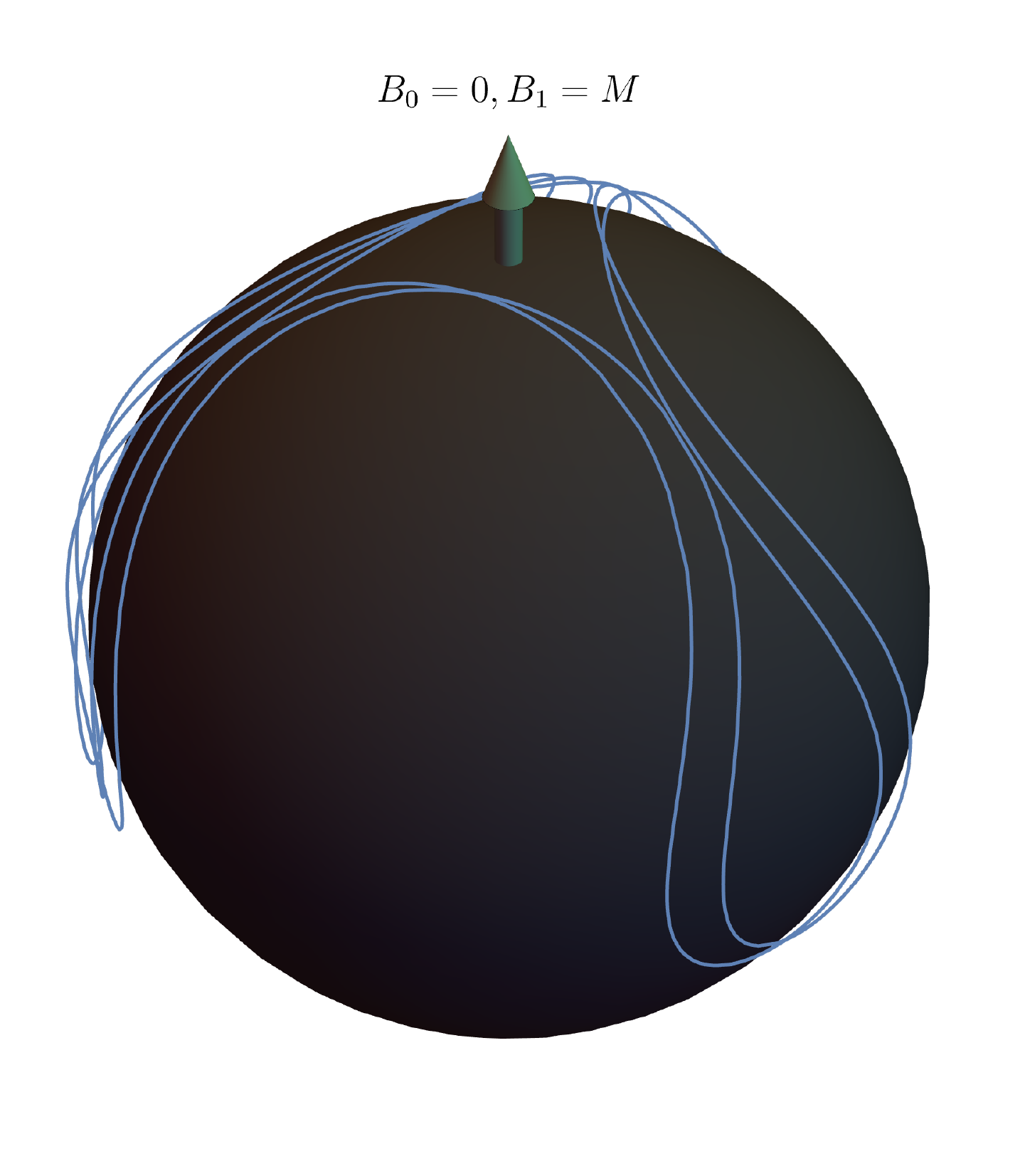}
   \newline
    \includegraphics[
    width=0.4\textwidth]{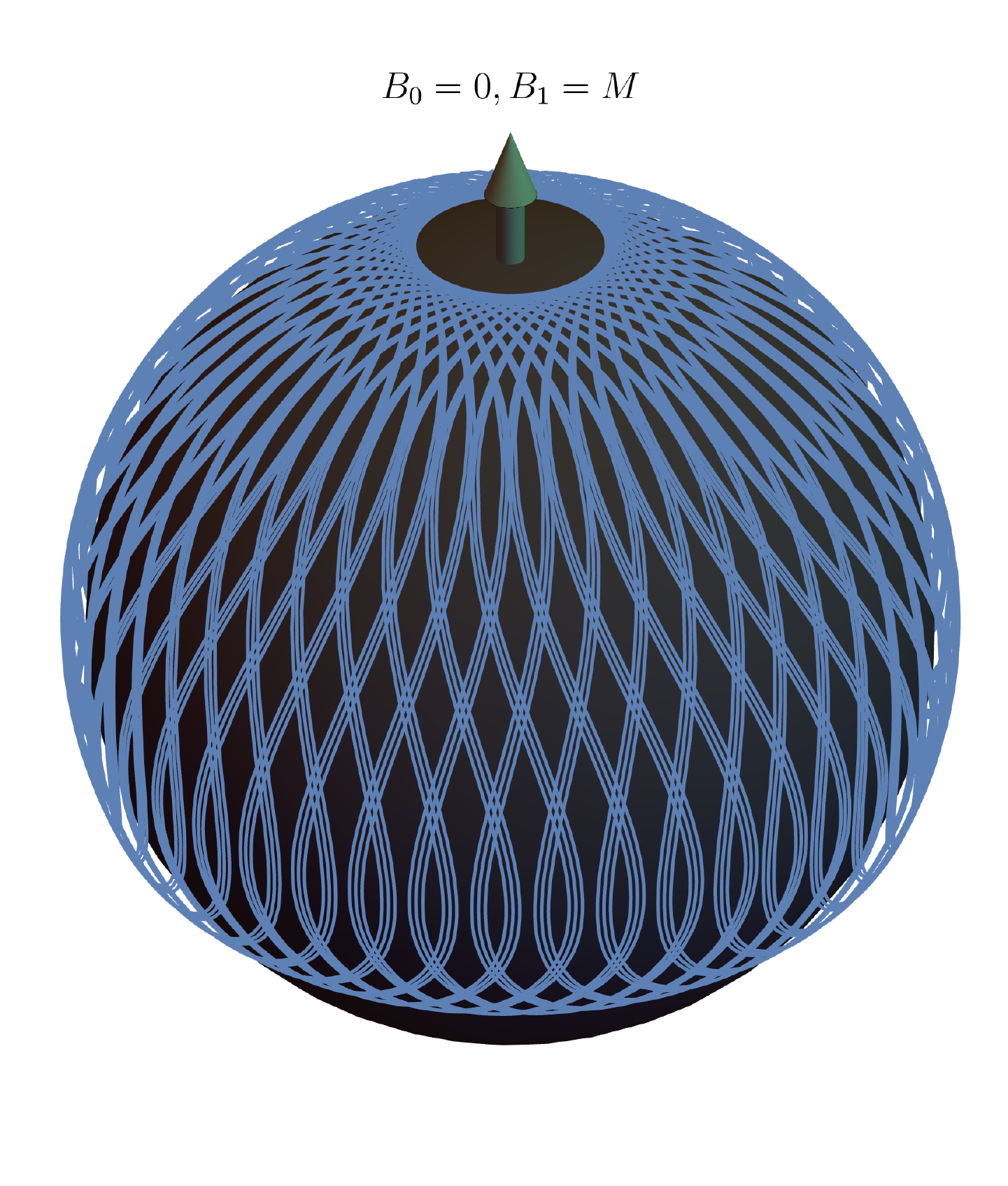}
\caption{{\bf Carrollian migration on the horizon.} We display the Carrollian motion reminiscent of the monarch butterfly migration for the choice of $\theta_0=\pi/2=\varphi_0$ for $B_0=M, B_1=0$, and $r_+=\frac{8}{5}M, a=\frac{4}{5}M$. The two figures show the same motion of one particle; the only difference is the number of revolutions displayed.} 
\label{fig:migration}
\end{figure}

{

A consequence of this precession is that the motion of Carrollian particles 
{does not simply correspond to the integral lines of the vector fields} seen in Fig~\ref{fig:vel1}, and {the non-trivial dependence on $v$ allows} for more interesting trajectories. {To illustrate this explicitly, 
we solve the} equation of motion \eqref{eq:EoM} numerically for different values of initial conditions 
\be 
\varphi(0)=\varphi_0\,,\quad \theta(0)=\theta_0\,.  
\ee
In Fig.~\ref{fig:migration} we display the migrating motion, where Carrollian particles move from regions located around the equator to regions around the poles. 
The observed motion of Carrollian particles seems integrable and very} similar to the motion of particles on a sphere rotating around a rotating axis, which is also not chaotic.

Finally, notice that although the curves in Fig.~\ref{fig:migration} overlap, it is not possible for Carrollian particles to collide with each other. This can be seen from the theorem of uniqueness and existence of solutions to ODEs and by noticing that the equations of motion of the Carrollian particle are of first order: A collision of two Carrollian particles would imply that two solutions would be possible starting from the event where the collision happens, which would contradict uniqueness. As far as we can conclude, Carrollian particles on a black hole horizon in an external magnetic field follow a very organized flow that approximately rotates around a precessing axis.

\section{Summary}\label{sec5}

{In this paper we have demonstrated that Carrollian particles can feature a rather non-trivial motion on the horizon of a black hole. To this purpose we have considered the Kerr black hole that is magnetized by a test asymptotically uniform magnetic field whose asymptotic axis of symmetry is tilted with respect to the rotational axis of the hole. As the horizon is equipped with a preferred Carrollian time, this gives rise to a `time dependent' magnetic field on the horizon and in consequence results in a (non-chaotic) `migratory' motion of Carrollian particles, which is   characterized by two vortices that precess around the black hole due to the inertial frame dragging on the horizon.  
} 


{
It remains to be seen, whether such motion of Carrollian particles (if realized in nature) would lead to some observable features, or whether the very existence of the Carrollian particles remains only an interesting theoretical possibility. 
}

\section{Acknowledgements}

J.B. 
is grateful for support from 
GA{\v C}R 21/112685S grant of 
the Czech Science Foundation, while
D.K. acknowledges GA{\v C}R 23-07457S grant  
of the Czech Science Foundation.
T.R.P. acknowledges support from the Natural Sciences and Engineering Research Council of Canada (NSERC) via the Vanier Canada Graduate Scholarship. Research at Perimeter Institute is supported in part by the Government of Canada through the Department of Innovation, Science and Industry Canada and by the Province of Ontario through the Ministry of Colleges and Universities.  {Perimeter Institute and the University of Waterloo are situated on the Haldimand Tract, land that was promised to the Haudenosaunee of the Six Nations of the Grand River, and is within the territory of the Neutral, Anishnawbe, and Haudenosaunee peoples.


\begin{thebibliography}{10}

\bibitem{Bergshoeff:2014jla}
E.~Bergshoeff, J.~Gomis and G.~Longhi, \emph{{Dynamics of Carroll Particles}},
  \href{https://doi.org/10.1088/0264-9381/31/20/205009}{\emph{Class. Quant.
  Grav.} {\bfseries 31} (2014) 205009}
  [\href{https://arxiv.org/abs/1405.2264}{{\ttfamily 1405.2264}}].

\bibitem{Freidel:2022bai}
L.~Freidel and P.~Jai-akson, \emph{{Carrollian hydrodynamics from symmetries}},
  \href{https://doi.org/10.1088/1361-6382/acb194}{\emph{Class. Quant. Grav.}
  {\bfseries 40} (2023) 055009}
  [\href{https://arxiv.org/abs/2209.03328}{{\ttfamily 2209.03328}}].

\bibitem{Freidel:2022vjq}
L.~Freidel and P.~Jai-akson, \emph{Carrollian hydrodynamics and symplectic
  structure on stretched horizons},  11, 2022.

\bibitem{Bagchi:2023ysc}
A.~Bagchi, K.S.~Kolekar and A.~Shukla, \emph{{Carrollian Origins of Bjorken
  Flow}},  \href{https://arxiv.org/abs/2302.03053}{{\ttfamily 2302.03053}}.

\bibitem{VD1966}
V.D.S.~Gupta, \emph{On an {A}nalogue of the {G}alileo {G}roup}, {\emph{Il Nuovo
  Cimento} {\bfseries 51} (1966) }.

\bibitem{Duval:2014uva}
C.~Duval, G.W.~Gibbons and P.A.~Horvathy, \emph{{Conformal Carroll groups and
  BMS symmetry}},
  \href{https://doi.org/10.1088/0264-9381/31/9/092001}{\emph{Class. Quant.
  Grav.} {\bfseries 31} (2014) 092001}
  [\href{https://arxiv.org/abs/1402.5894}{{\ttfamily 1402.5894}}].

\bibitem{Marsot:2022qkx}
L.~Marsot, P.-M.~Zhang and P.~Horvathy, \emph{{Anyonic spin-Hall effect on the
  black hole horizon}},
  \href{https://doi.org/10.1103/PhysRevD.106.L121503}{\emph{Phys. Rev. D}
  {\bfseries 106} (2022) L121503}
  [\href{https://arxiv.org/abs/2207.06302}{{\ttfamily 2207.06302}}].

\bibitem{Marsot:2022imf}
L.~Marsot, P.M.~Zhang, M.~Chernodub and P.A.~Horvathy, \emph{{Hall motions in
  Carroll dynamics}},  \href{https://arxiv.org/abs/2212.02360}{{\ttfamily
  2212.02360}}.

\bibitem{hirsch1999}
J.~Hirsch, \emph{{Spin Hall effect}}, {\emph{{Physical Review Letters}}
  {\bfseries 83} (1999) 1834}.

\bibitem{Harte:2022dpo}
A.I.~Harte and M.A.~Oancea, \emph{{Spin Hall effects and the localization of
  massless spinning particles}},
  \href{https://doi.org/10.1103/PhysRevD.105.104061}{\emph{Phys. Rev. D}
  {\bfseries 105} (2022) 104061}
  [\href{https://arxiv.org/abs/2203.01753}{{\ttfamily 2203.01753}}].

\bibitem{Gray:2022svz}
F.~Gray, D.~Kubiz{\v n}{\'a}k, T.R.~Perche and J.~Redondo-Yuste,
  \emph{{Carrollian Motion in Magnetized Black Hole Horizons}},
  \href{https://arxiv.org/abs/2211.13695}{{\ttfamily 2211.13695}}.

\bibitem{bivcak1976stationary}
J.~Bi{\v{c}}{\'a}k and L.~Dvo{\v{r}}{\'a}k, \emph{{Stationary electromagnetic
  fields around black holes. II. General solutions and the fields of some
  special sources near a Kerr black hole}}, {\emph{General Relativity and
  Gravitation} {\bfseries 7} (1976) 959}.

\bibitem{bivcak1985magnetic}
J.~Bi{\v{c}}{\'a}k and V.~Jani{\v{s}}, \emph{Magnetic fluxes across black
  holes}, {\emph{Mon. Not. Roy. Astron.
  Soc.} {\bfseries
  212} (1985) 899}.

\bibitem{hawking1972black}
S.W.~Hawking, \emph{Black holes in general relativity}, {\emph{Communications
  in Mathematical Physics} {\bfseries 25} (1972) 152}.

\bibitem{press1972time}
W.H.~Press, \emph{Time evolution of a rotating black hole immersed in a static
  scalar field}, {\emph{The Astrophysical Journal} {\bfseries 175} (1972) 243}.

\bibitem{thorne1986black}
K.S.~Thorne, R.H.~Price and D.A.~MacDonald, \emph{{Black Holes: The Membrane
  Paradigm}}, Yale university press (1986).

\bibitem{king1977magnetic}
A.~King and J.~Lasota, \emph{Magnetic alignment of rotating black holes and
  accretion discs.}, {\emph{Astronomy and Astrophysics} {\bfseries 58} (1977)
  175}.

\bibitem{Kim:2002ei}
H.~Kim, H.K.~Lee and C.H.~Lee, \emph{{Magnetic alignment process : A New
  mechanism for extracting energy from rotating black holes}},
  \href{https://doi.org/10.1088/1475-7516/2003/09/001}{\emph{JCAP} {\bfseries
  09} (2003) 001} [\href{https://arxiv.org/abs/astro-ph/0206171}{{\ttfamily
  astro-ph/0206171}}].

\bibitem{McKinney:2012wd}
J.C.~McKinney, A.~Tchekhovskoy and R.D.~Blandford, \emph{{Alignment of
  Magnetized Accretion Disks and Relativistic Jets with Spinning Black Holes}},
  \href{https://doi.org/10.1126/science.1230811}{\emph{Science} {\bfseries 339}
  (2013) 49} [\href{https://arxiv.org/abs/1211.3651}{{\ttfamily 1211.3651}}].

\bibitem{Liska:2018ayk}
M.~Liska, A.~Tchekhovskoy, A.~Ingram and M.~van~der Klis,
  \emph{{Bardeen\textendash{}Petterson alignment, jets, and magnetic truncation
  in GRMHD simulations of tilted thin accretion discs}},
  \href{https://doi.org/10.1093/mnras/stz834}{\emph{Mon. Not. Roy. Astron.
  Soc.} {\bfseries 487} (2019) 550}
  [\href{https://arxiv.org/abs/1810.00883}{{\ttfamily 1810.00883}}].

\bibitem{reppert2018demystifying}
S.M.~Reppert and J.C.~de~Roode, \emph{Demystifying monarch butterfly
  migration}, {\emph{Current Biology} {\bfseries 28} (2018) R1009}.

\bibitem{Chamblin:1998qm}
A.~Chamblin, R.~Emparan and G.W.~Gibbons, \emph{{Superconducting $p$-branes and
  extremal black holes}},
  \href{https://doi.org/10.1103/PhysRevD.58.084009}{\emph{Phys. Rev. D}
  {\bfseries 58} (1998) 084009}
  [\href{https://arxiv.org/abs/hep-th/9806017}{{\ttfamily hep-th/9806017}}].

\bibitem{Penna:2014aza}
R.F.~Penna, \emph{{Black hole Meissner effect and Blandford-Znajek jets}},
  \href{https://doi.org/10.1103/PhysRevD.89.104057}{\emph{Phys. Rev. D}
  {\bfseries 89} (2014) 104057}
  [\href{https://arxiv.org/abs/1403.0938}{{\ttfamily 1403.0938}}].

\bibitem{Bicak:2015lxa}
J.~Bi\v{c}\'ak and F.~Hejda, \emph{{Near-horizon description of extremal
  magnetized stationary black holes and Meissner effect}},
  \href{https://doi.org/10.1103/PhysRevD.92.104006}{\emph{Phys. Rev. D}
  {\bfseries 92} (2015) 104006}
  [\href{https://arxiv.org/abs/1510.01911}{{\ttfamily 1510.01911}}].

\bibitem{Gregory:2013xca}
R.~Gregory, D.~Kubiz{\v n}{\'a}k and D.~Wills, \emph{{Rotating black hole
  hair}}, \href{https://doi.org/10.1007/JHEP06(2013)023}{\emph{JHEP} {\bfseries
  06} (2013) 023} [\href{https://arxiv.org/abs/1303.0519}{{\ttfamily
  1303.0519}}].

\bibitem{Karas:2012mp}
V.~Karas, O.~Kop{\'a}{\v c}ek and D.~Kunneriath, \emph{{Influence of
  frame-dragging on magnetic null points near rotating black hole}},
  \href{https://doi.org/10.1088/0264-9381/29/3/035010}{\emph{Class. Quant.
  Grav.} {\bfseries 29} (2012) 035010}
  [\href{https://arxiv.org/abs/1201.0009}{{\ttfamily 1201.0009}}].

\bibitem{Donnay:2019jiz}
L.~Donnay and C.~Marteau, \emph{{Carrollian Physics at the Black Hole
  Horizon}}, \href{https://doi.org/10.1088/1361-6382/ab2fd5}{\emph{Class.
  Quant. Grav.} {\bfseries 36} (2019) 165002}
  [\href{https://arxiv.org/abs/1903.09654}{{\ttfamily 1903.09654}}].

\bibitem{Bergshoeff:2022eog}
E.~Bergshoeff, J.~Figueroa-O'Farrill and J.~Gomis, \emph{{A non-lorentzian
  primer}}, {\emph{SciPost Physics Lecture Notes}}, submitted, 2022.

\bibitem{Frolov:2006yb}
V.P.~Frolov, \emph{{Embedding of the Kerr-Newman black hole surface in
  Euclidean space}},
  \href{https://doi.org/10.1103/PhysRevD.73.064021}{\emph{Phys. Rev. D}
  {\bfseries 73} (2006) 064021}
  [\href{https://arxiv.org/abs/gr-qc/0601104}{{\ttfamily gr-qc/0601104}}].

\end{thebibliography}

\providecommand{\href}[2]{#2}\begingroup\raggedright\endgroup

\end{document}